\documentstyle[twocolumn,seceq,epsf]{jpsj}
\def\runtitle{Charge and Orbital Ordering in Bi$_x$V$_8$O$_{16}$}
\def\runauthor{Yoshinori {\sc Shibata} and Yukinori {\sc Ohta}}
\title{Charge and Orbital Ordering in the Triangular-Lattice 
$t_{2g}$-Orbital System in One Dimension:\\
A Possible Ground State of Bi$_x$V$_8$O$_{16}$}
\author{ Yoshinori {\sc Shibata}$^{1}$
\footnote{E-mail: shibata@physics.s.chiba-u.ac.jp}
and Yukinori {\sc Ohta}$^{1,2}$
\footnote{E-mail: ohta@science.s.chiba-u.ac.jp}}
\inst{$^1$Graduate School of Science and Technology, Chiba University, 
Chiba 263-8522\\
$^2$Department of Physics, Chiba University, Chiba 263-8522}
\recdate{September 25, 2001}
\abst{
We consider the possible charge and orbital ordering in a Hollandite 
compound Bi$_x$V$_8$O$_{16}$, which is a new one-dimensional 
triangular-lattice $t_{2g}$-orbital system.  Using the strong-coupling 
perturbation theory, we derive the effective spin-orbit Hamiltonian 
in the approximation neglecting the small off-diagonal hopping 
parameters or orbital fluctuation, whereby we obtain the spin Hamiltonians 
in the partial space of each orbital-ordering pattern.  We then apply 
an exact-diagonalization technique on small clusters to these spin 
Hamiltonians and calculate the ground-state phase diagram.  We find 
that a variety of orbital-ordering patterns appear in the parameter 
space, which include the state characterized by the partial singlet 
formation consistent with recent NMR experiment.}
\kword{charge ordering, orbital ordering, vanadate, Hollandite, 
Bi$_x$V$_8$O$_{16}$, perturbation theory}
\begin{document}
\sloppy
\maketitle

\section{Introduction}

Orbital physics in transition-metal oxides, such as the orbital 
ordering (OO) of the $e_g$-spins of Mn-oxides, has attracted much 
attention in the research field of strongly correlated electron 
systems.\cite{tokura}  
Recently, Kato {\it et al.} \cite{kato} measured the magnetic 
susceptibility and resistivity of Bi$_x$V$_8$O$_{16}$ with 
$1.72<x<1.8$ and found a metal-insulator transition (MIT) at 
$T<80$ K.  The crystal structure of this compound belongs to 
a group of Hollandite-type phases and has a V$_8$O$_{16}$ 
framework composed of double strings of edge-shared VO$_6$ 
octahedra as shown in Fig.~1.  Kato {\it et al.}\cite{kato} 
suggested a mechanism of the transition that the charge 
ordering (CO) between V$^{3+}$ and V$^{4+}$ accompanied by 
an OO occurs with MIT.  
%%%% FIG.1 %%%%
\begin{figure}[h]
\vspace{20pt}
\begin{center}
\epsfxsize=6.0cm \epsfbox{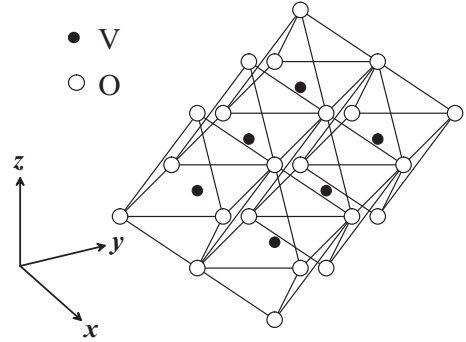}
\caption{Schematic representation of the double string of 
edge-shared VO$_{6}$ octahedra in Bi$_x$V$_8$O$_{16}$.}
\end{center}
\label{fig:1}
\end{figure}
%%%%%%%%%%%%%%%

In the viewpoint of orbital physics, this system may be regarded 
as a one-dimensional (1D) version of LiVO$_2$ known as a possible 
OO system of $t_{2g}$-orbitals on the 2D triangular lattice of 
$S=1$ spins,\cite{pen1,pen2} although our system Bi$_x$V$_8$O$_{16}$ 
has the average valence of V$^{(3+1/3)+}$ at $x=16/9$ and thus 
is in the mixed valent state of 
V$^{3+}$ : V$^{4+}$ = $3d^2$ : $3d^1$ = 2 : 1.\cite{kato} 
The central issue in the present system is therefore the mechanism 
of the MIT concerning how the highly frustrated spin, charge, and 
orbital degrees of freedom at high temperatures are relaxed by 
lowering temperatures and what type of the ground state is 
realized at zero temperature.  

To consider this issue, we study in this paper the possible ground 
state of this system.  We thereby assume that the V-ions, which have 
the triply degenerate $t_{2g}$-orbitals with intra- and inter-orbital 
Coulomb interactions as well as Hund's rule coupling, form the 1D 
triangular lattice and that the intersite Coulomb repulsions are 
strong enough for the electrons to be localized to form a spatial 
CO pattern.  
We then use the strong-coupling perturbation theory \cite{kugel} to 
derive the effective spin-orbit Hamiltonian, which turns out to be 
block-diagonal with vanishing orbital-off-diagonal elements within 
the approximation used.  	
We employ a numerical exact-diagonalization technique on small 
clusters of this Hamiltonian to consider the possible OO spatial 
patterns in the ground state.  
We find that a variety of the ground-state phases appear in the 
parameter space.  We also discuss the spin degrees of freedom of the 
obtained OO patterns.  We argue that the state characterized by 
the singlet formation of two $S=1$ spins on the neighboring 
V$^{3+}$-ions with remaining nearly-free $S=1/2$ spins on the 
V$^{4+}$-ions might be relevant in the present material.  

Although the experimental data on this new system Bi$_x$V$_8$O$_{16}$ 
are quite limited at present, we hope that our first theoretical study 
on its charge, orbital, and spin degrees of freedom would stimulate 
further experimental studies of this intriguing material.  
%%%% FIG.2 %%%%
\begin{figure}[b]
\vspace{25pt}
\begin{center}
\epsfxsize=4.2cm \epsfbox{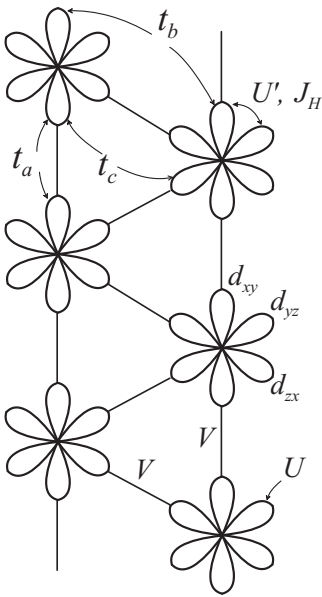}
\caption{Schematic representation of the $t_{2g}$-orbitals on 
the 1D triangular lattice.  Two of the four lobes for each 
of the three $t_{2g}$-orbitals are drawn.}
\end{center}
\label{fig:2}
\end{figure}
%%%%%%%%%%%%%%%

\section{Model}

Our starting high-energy Hamiltonian is of the following form: 
\begin{eqnarray*}
&&H=H_0+H_t\\
&&H_0=V\sum_{\langle ij\rangle}n_in_j
-J_H\sum_{i\sigma\sigma',\alpha\neq\beta}c^\dagger_{i\alpha\sigma}
c^\dagger_{i\beta\sigma'}c_{i\beta\sigma}c_{i\alpha\sigma'}\\
&&~~~~~~~~+U'\sum_{i,\alpha\neq\beta}n_{i\alpha}n_{i\beta}
+U\sum_{i\alpha}n_{i\alpha\uparrow}n_{i\alpha\downarrow}\\
&&H_t=-\sum_{\langle i\alpha,j\beta \rangle,\sigma}
t_{i\alpha,j\beta}(c^\dagger_{i\alpha\sigma}c_{j\beta\sigma}
+{\rm H.c.})
\end{eqnarray*}
where $V$ is the intersite Coulomb repulsion, $J_H$ is the Hund's 
rule coupling, and $U$ and $U'$ are the intra- and inter-orbital 
Coulomb repulsions, respectively.  $\langle\cdots\rangle$ stands 
for the nearest-neighbor pair of sites.  
We neglect the crystal-field splittings among the $t_{2g}$-orbitals 
for simplicity because the gain in kinetic energy may be much 
larger than the splittings as has been assumed in ref.\cite{pen1}  
We also assume the relation $U'=U-2J_H$, which is valid in the atomic 
limit.\cite{brandow}  $t_{i\alpha,j\beta}$ is the hopping parameter 
between the orbital $\alpha$ on site $i$ and orbital $\beta$ on site 
$j$ where $\alpha,\beta \in \{d_{xy},d_{yz},d_{zx}\}$ in the 
coordinate system shown in Fig.~1.  We retain the direct hoppings 
between the $t_{2g}$-orbitals on the V-ions because the indirect 
hoppings via the O-ions are rather small.\cite{pen1}  
Independent nearest-neighbor $t_{i\alpha,j\beta}$ parameters are 
$t_a$, $t_b$, and $t_c$ as shown in Fig.~2.  
$c_{i\alpha\sigma}^\dagger$ $(c_{i\alpha\sigma})$ is the electron 
creation (annihilation) operator at site $i$, orbital $\alpha$, 
and spin $\sigma$, and 
$n_{i\alpha\sigma}=c^\dagger_{i\alpha\sigma}c_{i\alpha\sigma}$ 
is the number operator.  We also define 
$n_{i\alpha}=n_{i\alpha\uparrow}+n_{i\alpha\downarrow}$ with 
$\sigma=\uparrow,\downarrow$ and $n_i=\sum_\alpha n_{i\alpha}$.  

\section{Charge ordering}

First, we consider the CO in the ground state.  We assume the 
limit of no doubly occupied orbitals.  
Then, if we assume that the inter-site Coulomb repulsions $V$ 
(as well as the inter-orbital Coulomb repulsion $U'$) is much 
larger than the hopping parameters, we find the ground state 
of the system to be charge ordered.  
It is readily noticed that the lowest-energy CO state has the 
ordered pattern like $\cdots V^{3+}V^{3+}V^{4+}\cdots$ as shown 
in Fig.~3.  At $H_t=0$, this state has the ground-state energy 
\begin{eqnarray*}
E_0=\frac{N}{6}(4U'-4J_H+32V)
\end{eqnarray*}
when there are no doubly occupied orbitals, whereas the states 
containing V$^{2+}$ or V$^{5+}$ have the ground-state energy 
\begin{eqnarray*}
E_0=\frac{N}{6}(6U'-6J_H+28V)
\end{eqnarray*}
for $\cdots{\rm V}^{2+}{\rm V}^{4+}{\rm V}^{4+}\cdots$, 
and 
\begin{eqnarray*}
E_0=\frac{N}{6}(8U'-8J_H+24V)
\end{eqnarray*}
for $\cdots{\rm V}^{2+}{\rm V}^{3+}{\rm V}^{5+}\cdots$.  
Since the states containing V$^{2+}$ or V$^{5+}$ are highly 
unrealistic, we should have the condition 
\begin{eqnarray*}
U'-2V-J_H>0
\end{eqnarray*}
for the presence of CO of only V$^{3+}$ and V$^{4+}$ ions.  
%%%% FIG.3 %%%%
\begin{figure}[h]
\vspace{25pt}
\begin{center}
\epsfxsize=6.0cm \epsfbox{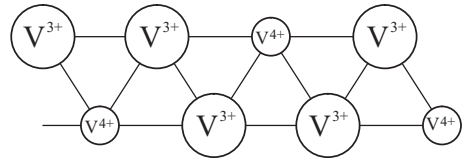}
\caption{Schematic representation of the CO pattern in 
the 1D triangular lattice.}
\end{center}
\label{fig:3}
\end{figure}
%%%%%%%%%%%%%%%

The ground state of the unperturbed ($H_t=0$) Hamiltonian is then 
of the degeneracy 
\begin{eqnarray*}
M=3^N\cdot 3^{2N/3}\cdot 2^{N/3}
\end{eqnarray*}
because there are $3^N$ choices of orbitals, $3^{2N/3}$ choices 
of $S=1$ spins, and $2^{N/3}$ choices of $S=1/2$ spins, where $N$ 
is the number of sites in the system.  The degeneracy is lifted 
by the perturbation of small hopping parameters $t_{i\alpha,j\beta}$ 
as shown in the next section.  

\section{Effective Hamiltonian}

We carry out the second-order perturbation calculation with 
respect to $t_{i\alpha,j\beta}$ assuming that $t_{i\alpha,j\beta}$ 
is much smaller than $U$, $U'$, and $V$.  The degeneracy is 
lifted by the perturbation and the effective spin-orbit Hamiltonian 
is obtained as follows: 
\begin{eqnarray*}
H_{\rm eff}=H_0-\sum_{\mu\mu'}|\mu\rangle\sum_n
{{\langle\mu|H_t|n\rangle\langle n|H_t|\mu'\rangle}\over{E_n-E_0}}
\langle\mu'|
\end{eqnarray*}
where $|\mu\rangle$ and $|\mu\rangle'$ $(\mu,\mu'=1,\cdots,M)$ are 
the $M$ independent eigenvectors of the ground state of $H_0$ and 
$|n\rangle$ are the $n$-th excited states of $H_0$.  $E_0$ and $E_n$ 
are the corresponding eigenenegies of $H_0$.  

To find the expressions for the effective Hamiltonian, let us first 
prepare the degenerate eigenstates of $H_0$: 
\begin{eqnarray*}
|\psi_0\rangle=\sum_{\{q\}}C(\{q\})
\prod_{i}|s_i,s^z_i,n_i,n^d_i\rangle_{\{\alpha_i\}}
\end{eqnarray*}
where $s_i$ and $s_i^z$ are the spin quantum numbers at site-$i$, 
$n_i$ is the number of electrons at site-$i$, $n^d_i$ is the 
number of doubly occupied orbitals at site-$i$, and $\{\alpha_i\}$ 
is the set of occupied orbitals.  $\{q\}$ represents a set of all 
these qantum numbers.  
The basis states $|s_i,s^z_i,n_i,n^d_i\rangle_{\{\alpha_i\}}$ may 
be of the following form: 
\\
\begin{description}
\item{(i)} For the $d^1$-site, we have 
\begin{eqnarray*}
&&|1/2,1/2,1,0\rangle_{\alpha}=|\uparrow\rangle_{\alpha}\\
&&|1/2,-1/2,1,0\rangle_{\alpha}=|\downarrow\rangle_{\alpha}
\end{eqnarray*}
with $S=1/2$ and $\alpha=xy,\,yz,\,zx$.  
%\\
\item{(ii)} For the $d^2$-site, we have 
\begin{eqnarray*}
&&|1,1,2,0\rangle_{\alpha\beta}
=|\uparrow \rangle_{\alpha}|\uparrow \rangle_{\beta}\\
&&|1,0,2,0\rangle_{\alpha\beta}
=\frac{1}{\sqrt{2}}
(|\uparrow \rangle_{\alpha}|\downarrow\rangle_{\beta}
+|\downarrow \rangle_{\alpha}|\uparrow\rangle_{\beta})\\
&&|1,-1,2,0\rangle_{\alpha\beta}
=|\downarrow \rangle_{\alpha}|\downarrow\rangle_{\beta}
\end{eqnarray*}
for $S=1$, and 
\begin{eqnarray*}
&&|0,0,2,0\rangle_{\alpha\beta}
=\frac{1}{\sqrt{2}}
(|\uparrow \rangle_{\alpha}|\downarrow\rangle_{\beta}
-|\downarrow \rangle_{\alpha}|\uparrow \rangle_{\beta})\\
&&|0,0,2,1\rangle_{\alpha}=|d\rangle_{\alpha}
\end{eqnarray*}
for $S=0$, where $d$ means the double occupancy.  
%\\
\item{(iii)} For the $d^3$-site, we have 
\begin{eqnarray*}
&&|3/2,3/2,3,0\rangle_{\alpha\beta\gamma}
=|\uparrow\uparrow\uparrow\rangle\\
&&|3/2,1/2,3,0\rangle_{\alpha\beta\gamma}
=\frac{1}{\sqrt{3}}(|\uparrow\uparrow\downarrow\rangle
+|\uparrow\downarrow\uparrow\rangle
+|\downarrow\uparrow\uparrow\rangle\\
&&|3/2,-1/2,3,0\rangle_{\alpha\beta\gamma}
=\frac{1}{\sqrt{3}}(|\downarrow\downarrow\uparrow\rangle
+|\downarrow\uparrow\downarrow\rangle
+|\uparrow\downarrow\downarrow\rangle\\
&&|3/2,-3/2,3,0\rangle_{\alpha\beta\gamma}
=|\downarrow\downarrow\downarrow\rangle
\end{eqnarray*}
for $S=3/2$, and
\begin{eqnarray*}
&&|1/2,1/2,3,0\rangle^{(1)}_{\alpha\beta\gamma}
=\frac{1}{\sqrt{2}}(|\uparrow\downarrow\uparrow\rangle
-|\downarrow\uparrow\uparrow\rangle)\\
&&|1/2,1/2,3,0\rangle^{(2)}_{\alpha\beta\gamma}
=\frac{1}{\sqrt{5}}(|\uparrow\downarrow\uparrow\rangle
+|\downarrow\uparrow\uparrow\rangle
-2|\uparrow\uparrow\downarrow\rangle)\\
&&|1/2,1/2,3,1\rangle_{\alpha\beta}
=|d\rangle_{\alpha}|\uparrow\rangle_{\beta}\\
&&|1/2,-1/2,3,0\rangle^{(1)}_{\alpha\beta\gamma}
=\frac{1}{\sqrt{2}}(|\uparrow\downarrow\downarrow\rangle
-|\downarrow\uparrow\downarrow\rangle)\\
&&|1/2,-1/2,3,0\rangle^{(2)}_{\alpha\beta\gamma}
=\frac{1}{\sqrt{5}}(|\uparrow\downarrow\downarrow\rangle
+|\downarrow\uparrow\downarrow\rangle
-2|\downarrow\downarrow\uparrow\rangle)\\
&&|1/2,-1/2,3,1\rangle_{\alpha\beta}
=|d\rangle_{\alpha}|\downarrow\rangle_{\beta}
\end{eqnarray*}
\end{description}
for $S=1/2$, where we use the notation 
$|\sigma_1\sigma_2\sigma_3\rangle
=|\sigma_1\rangle_\alpha|\sigma_2\rangle_\beta|\sigma_3\rangle_\gamma$.  
These states are used in the following as the intermediate states 
of the second-order perturbation processes.  

We here introduce an approximation;  because the hopping 
parameters $t_{i\alpha,j\beta}$ take the values 
$t_a\gg t_b\simeq t_c$ \cite{pen1}, we assume $t_b=t_c=0$ 
for simplicity as in ref.\cite{pen1}.  This means that the 
terms like $H_{ij}^{\rm eff}\propto t_at_c$ and 
$H_{ij}^{\rm eff}\propto t_bt_c$ are all neglected, 
retaining only the terms like $H_{ij}^{\rm eff}\propto t_a^2$ 
in the second-order processes.  We note that the orbital 
fluctuations are completely suppressed in this approximation 
because only two orbitals connected with the {\it diagonal} 
hopping $t_a$ comes out and no {\it off-diagonal} hopping 
terms appear in the effectve Hamiltonian.  In other words, we 
obtain the effective spin-orbit Hamiltonian consisting of 
orbital-diagonal spin-sub-blocks with vanishing orbital 
off-diagonal blocks.  
%%%% FIG.4 %%%%
\begin{figure}[h]
\vspace{25pt}
\begin{center}
\epsfxsize=5.0cm \epsfbox{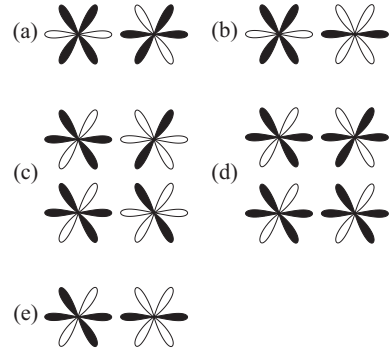}
\caption{Schematic representation of five types of the 
bonds with different exchange interactions.  
(a) FM-1: the ferromagnetic bond with the process 
$d_i^2d_j^2 \rightarrow d_i^3d_j^1 \rightarrow d_i^2d_j^2$, 
(b) FM-2: the ferromagnetic bond with the process 
$d_i^2d_j^1 \rightarrow d_i^3d_j^0 \rightarrow d_i^2d_j^1$, 
(c) FM-3: the ferromagnetic bond with the process 
$d_i^2d_j^1 \rightarrow d_i^1d_j^2 \rightarrow d_i^2d_j^1$, 
(d) AF-1: the antiferromagnetic bond with the process 
$d_i^2d_j^2 \rightarrow d_i^3d_j^1 ~{\rm or}~ d_i^1d_j^3 
\rightarrow d_i^2d_j^2$, and 
(e) AF-2: the antiferromagnetic bond with the process 
$d_i^2d_j^1 \rightarrow d_i^3d_j^0 ~{\rm or}~ d_i^1d_j^2 
\rightarrow d_i^2d_j^1$.}
\end{center}
\label{fig:4}
\end{figure}
%%%%%%%%%%%%%%%

We then have five types of bonds of spin exchange interactions 
as shown in Fig.~4; three of them are the bonds with ferromagnetic 
(FM) exchange interaction due to double-exchange mechanism and 
two of them are the bonds with antiferromagnetic (AF) exchange 
interaction due to kinetic-exchange mechanism.  Defining the 
spin-1 operator as ${\bf S}_i$ and spin-1/2 operator as 
${\bf s}_i$, we have the following Hamiltonian for each bond 
shown in Fig.~4.  
\\
\begin{description}
\item{(i)} The bond FM-1 with the process 
$d_i^2d_j^2\rightarrow d_i^3d_j^1\rightarrow d_i^2d_j^2$:
\begin{eqnarray*}
&&H_{ij}^{\rm eff}=-J{\bf S}_i\cdot{\bf S}_j+c\hat{1}\\
&&J=\frac{t_a^2}{3(U'-V-J_H)}-\frac{2t_a^2}{5(U'-V+2J_H)}\\
&&c=-\frac{2t_a^2}{3(U'-V-J_H)}-\frac{2t_a^2}{5(U'-V+2J_H)}
\end{eqnarray*}
where $\hat{1}$ is the unit operator.  
%\\
\item{(ii)} The bond FM-2 with the process 
$d_i^2d_j^1\rightarrow d_i^3d_j^0\rightarrow d_i^2d_j^1$:
\begin{eqnarray*}
&&H_{ij}^{\rm eff}=-2J{\bf S}_i\cdot{\bf s}_j+c\hat{1}\\
&&J=\frac{t_a^2}{3(2U'-3V-2J_H)}-\frac{2t_a^2}{5(2U'-3V+J_H)}\\
&&c=-\frac{2t_a^2}{3(2U'-3V-2J_H)}-\frac{2t_a^2}{5(2U'-3V+J_H)}
\end{eqnarray*}
%\\
\item{(iii)} The bond FM-3 with the process 
$d_i^2d_j^1\rightarrow d_i^1d_j^2\rightarrow d_i^2d_j^1$:
\begin{eqnarray*}
&&H_{ij}^{\rm eff}=-2J{\bf S}_i\cdot{\bf s}_j+c\hat{1}\\
&&J=\frac{t_a^2}{4V}-\frac{t_a^2}{4(V+2J_H)}\\
&&c=-\frac{3t_a^2}{4V}-\frac{t_a^2}{4(V+2J_H)}
\end{eqnarray*}
In the above three, we have the competing exchange couplings 
with positive and negative sign, but the sum is always positive 
for realistic values of the parameters $J_H$, $V$, and $U'$.  
%\\
\item{(iv)} The bond AF-1 with the process 
$d_i^2d_j^2\rightarrow d_i^3d_j^1$ or $d_i^1d_j^3\rightarrow d_i^2d_j^2$: 
\begin{eqnarray*}
&&H_{ij}^{\rm eff}=J{\bf S}_i\cdot{\bf S}_j+c\hat{1}\\
&&J=-c=\frac{t_a^2}{U-V+J_H}
\end{eqnarray*}
%\\
\item{(v)} The bond AF-2 with the process 
$d_i^2d_j^1\rightarrow d_i^3d_j^0$ or $d_i^1d_j^2\rightarrow d_i^2d_j^1$: 
\begin{eqnarray*}
&&H_{ij}^{\rm eff}=2J{\bf S}_i\cdot{\bf s}_j+c\hat{1}\\
&&J=-c=\frac{t_a^2}{2(U+U'-3V)}+\frac{t_a^2}{2(U-U'+V+J_H)}
\end{eqnarray*}
\end{description}
These two exchange couplings have the doubly occupied states in 
their intermediate state and thus they are antiferromagnetic.  

Note that the intersite Coulomb repulsion $V$ can be included 
in the bond Hamiltonian as shown above because, due to the fixed 
CO pattern of Fig.~3, all the intermediate states have the same 
intersite Coulomb energy, irrespective of the location of the 
bond.  Also noted is that the so-called three-site terms in 
the perturbation do not appear for the CO pattern to be fixed.  

Thus, we have the effective spin-orbit Hamiltonian 
\begin{eqnarray*}
H_{\rm eff}=\sum_{<ij>}H_{ij}^{\rm eff}
\end{eqnarray*}
where the sum runs over all the nearest-neighbor pairs of 
sites.  
Note that this effective spin-orbit Hamiltonian has the form 
of block-diagonal in the spin$\otimes$orbit space; i.e., 
orbital off-diagonal blocks are all zero.  In other words, we 
have $3^N$ OO patterns for the $N$-site systems, and for each 
of them, we have the spin Hamiltonian.  If we diagonalize all 
the spin Hamiltonians, we can obtain the eigenstates of our 
effective spin-orbit Hamiltonian.  
We use the Lanczos diagonalization technique on small clusters 
to diagonalize the spin sub-block Hamiltonians and by comparing 
the lowest enrgies obtained we find the ground-state OO 
patterns in the parameter space.  
%%%% TABLE I %%%%
\begin{table}[h]
\caption{Number of bonds in each OO pattern calculated for the 
12-site cluster.}
\label{table:1}
\begin{center}
\begin{tabular}{@{\hspace{\tabcolsep}\extracolsep{\fill}}ccccccc}
\hline
type of bond & phase-I & phase-II & phase-III & phase-IV \cr
\hline
FM-1          &   4     &    2     &    0      &    6     \cr
FM-2          &   4     &    2     &    0      &    4     \cr
FM-3          &  12     &   12     &    12     &    10    \cr
AF-1          &   4     &    4     &    4      &    2     \cr
AF-2          &   0     &    2     &    4      &    2     \cr
\hline
\end{tabular}
\end {center}
\end{table}
%%%%%%%%%%%%%%%%%
%%%% FIG.5 %%%%
\begin{figure}[h]
\vspace{25pt}
\begin{center}
\epsfysize=5.0cm \epsfbox{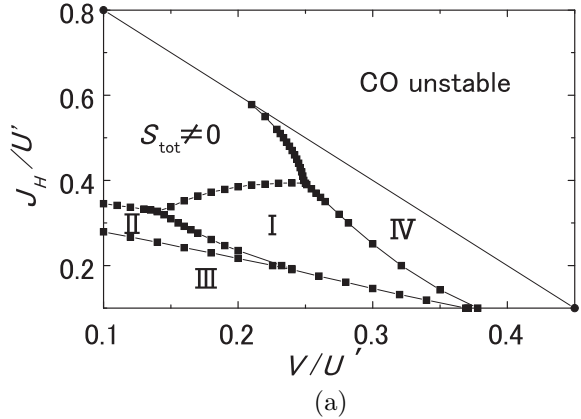}

(a)

\epsfysize=6.0cm \epsfbox{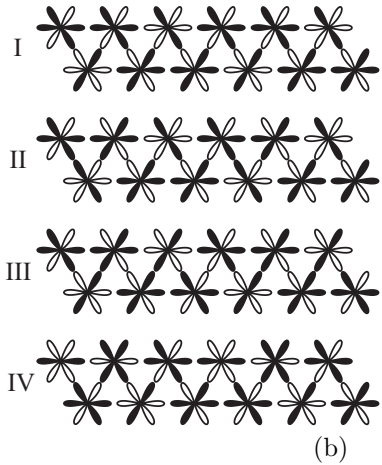}

(b)

\caption{(a) Calculated ground-state phase diagram of the effective 
spin-orbit Hamiltonian.  The phases I--IV have the OO spatial patterns 
shown in (b).  
The region indicated as $S_{\rm tot}>0$ has the OO pattern I but 
has higher total spins.  
The CO of V$^{3+}$ and V$^{4+}$ is unstable in the upper-right part 
of the phase diagram.  
(b) Calculated OO spatial patterns of the phases I--IV.  Shaded lobes 
have the electrons.}
\end{center}
\label{fig:5}
\end{figure}
%%%%%%%%%%%%%%%

\section{Phase diagram}

Now, let us calculate the ground-state phase diagram of the 
effective Hamiltonian derived in the previous section.  
We use the 12-site (corresponding to 36-orbital) cluster with 
eight $S=1$ spins and four $S=1/2$ spins (corresponding to the 
filling of 20 electrons) coupled with the derived exchange 
interactions in the lattice of the CO pattern; we thereby 
calculated the ground state of each spin Hamiltonian.  
The periodic boundary condition is used.  The calculated 
results for all possible OO patterns (where we assume the 
unit cell containing 6 sites) indicate that a variety of 
the OO patterns appear as the ground state, depending on 
the parameter values of $J_H/U'$ and $V/U'$ as shown in 
Fig.~5(a).  

The OO spatial patterns of the phases I--IV are illustrated in 
Fig.~5(b).  
Careful inspection of the patterns indicate that the gain in 
kinetic energy by the process FM-3 is the most important to 
stabilize these phases.  This is evident in Table I, where 
the numbers of types of bonds existing in the phases I--IV 
are listed; i.e., the OO patterns I--IV are stabilized by 
maximizing the number of the bond FM-3 where the constant 
$c$-value is predominantly lower than others as shown in 
Fig.~6(a).  
%%%% FIG.6 %%%%
\begin{figure}[h]
\vspace{25pt}
\begin{center}
\epsfxsize=5.5cm \epsfbox{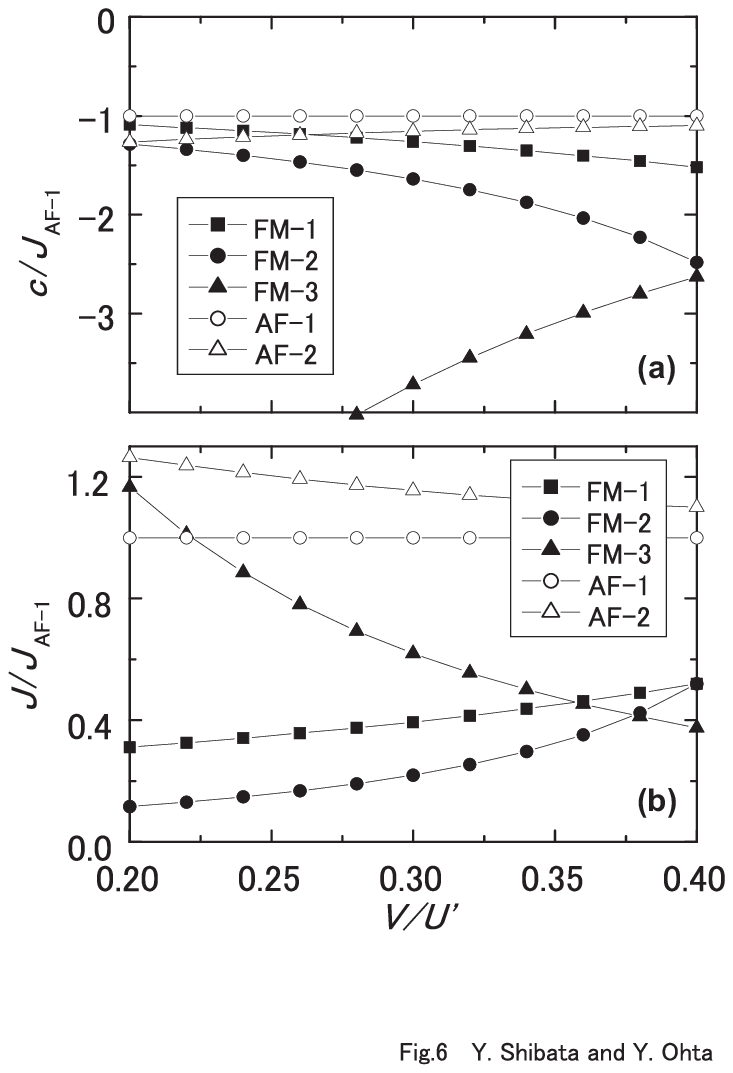}
\caption{Calculated values of (a) the exchange coupling 
constants $J$ and (b) constant terms $c$ in the effective 
spin Hamiltonians.  The values normalized by the exchange 
coupling constant of the bond AF-1 are plotted.}
\end{center}
\label{fig:6}
\end{figure}
%%%%%%%%%%%%%%%

We find in Fig.~6(b) that the exchange energies of the 
antiferromagnetic bonds are generally larger than the exchange 
energies of the ferromagnetic bonds; $J_{\rm AF}>J_{\rm F}$.  
This energy difference determines the detailed energy differences 
of the phases I--IV.  For example, it may be that the strong AF-1 
coupling promotes the singlet formation of the $S=1$ spin pair 
by lowering its energy, to result in the appearance of the 
relatively stable phase I.  
Also, when the value of $J_H$ is small, the strong antiferromagnetic 
couplings AF-1 and AF-2 stabilize the phase III as is noted 
in Table I.  On the other hand, when $J_H$ is large, the 
ferromagnetic couplings FM-1 and FM-2 stabilize the phase I 
as is also noted in Table I.  When $V/U'$ is small, the phase 
$S_{\rm tot}>0$, which has the same OO pattern as the phase I, 
becomes the ground state.  This may be due to the strong 
ferromagnetic coupling FM-3, which is rapidly enhanced when 
$V/U'$ becomes small as we find in Fig.~6(b).  

Because $J_H/U'\simeq 0.2$ in the real material but the value of 
$V/U'$ is not well-known, we may expect that the phases I--IV 
should all be possible to be realized.  However, the experimental 
data on the spin degrees of freedom may single out a possible 
phase as we will discuss in the next section.  

\section{Discussion}

Finally, let us discuss the spin degrees of freedom of the system.  
Speculated spin states of the phases I--IV are illustrated in 
Fig.~7, which is supported in part by our small-cluster calculations 
of the spin correlation functions for the spin Hamiltonians.  
We find that the variety of the OO patterns we have obtained result 
in a variety of the spin states as discussed below, although detailed 
numerical analyses of the Hamiltonians are needed for the definite 
descriptions of the states, which we leave for future study.  
%%%% FIG.7 %%%%
\begin{figure}[h]
\vspace{25pt}
\begin{center}
\epsfxsize=5.5cm \epsfbox{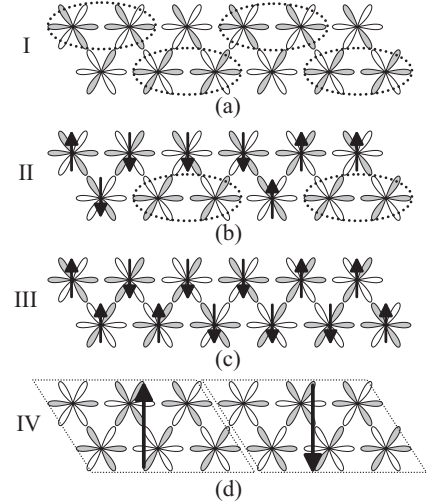}
\caption{Speculated spin states of the phases I--IV, where shaded 
lobes have the electrons, dotted circles represent the local 
spin-singlet state, and arrows indicate the spin direction.  
(a) Phase-I: Singlets of two $S=1$ spins and nearly-free $S=1/2$ 
spins coexist. 
(b) Phase-II: Intermediate state between phases I and III.  
(c) Phase-III: State without spin frustration and thus with strong 
AF spin correlation.  
(d) Phase-IV: AF state with local high-spin clusters.}
\end{center}
\label{fig:7}
\end{figure}
%%%%%%%%%%%%%%%

For phase I, we find that the OO spatial pattern may be consistent 
with the spin state where the spin structure may be of the 
{\it partial} singlet formation; i.e., 2/3 of V-ions have $S=1$ 
and form the spin singlet pairs, leaving $S=1/2$ free spins on 
1/3 of V-ions.  For phase III, we find that the state has no spin 
frustration and thus we have the strong local antiferromagnetic 
spin correlations.  For phase II, we have the intermediate spin 
state between the phases I and III.  For phase IV, we find that 
the state is formed by the local high-spin clusters of six V-ions, 
which are linked in the antiferromagnetic arrangement.  

We may then suggest the following picture for the charge and orbital 
structure of the ground state of Bi$_x$V$_8$O$_{16}$:  
(i) There occurs the CO where the V-ions order as 
$\cdots$V$^{3+}$V$^{3+}$V$^{4+}$$\cdots$ in the chain direction.  
Thus, one should observe the crystal structure of three-fold 
periodicity.  
(ii) The possible spatial patterns of orbitals selected among the 
three $t_{2g}$-orbitals are determined as the phase I in Fig.~5(b).  
(iii) The spin degrees of freedom are then considered to be the partial 
singlet formation.  This state may be consistent with the result of 
an NMR experiment recently made by Waki {\it et al.},\cite{waki} 
where it has been suggested that most of the spins form the singlet 
state with leaving a small amount of active spins, which then undergo 
an additional phase transition into the magnetic long-range order 
at lower temperatures.  

As for the MI phase transition of the real material, we suspect that 
the highly frustrated electronic state makes the system metallic 
at high temperatures, the CO occurs by lowering temperature to 
result in the MIT, and, triggered by this transition (but 
simultaneously), there occur the OO and spin-singlet formation.  
At lower temperatures, there appears the magnetic long-range 
order,\cite{kato} which may be due to the weak exchange coupling 
between the $S=1/2$ spins in our OO system (though not considered 
in this paper).  

The approximation used here neglects the orbital fluctuation 
completely.  The effect may however be important if we want to 
consider, e.g., the state at finite temperatures.  The use of 
the approximation may however be justified because we here want 
to answer the question ``what is the ground-state OO pattern 
if we assume the presence of OO''.  Possible absence of OO as 
in LaTiO$_3$ \cite{khaliullin} may of course be an interesting 
issue in the present system as well.  Jahn-Teller distortions 
as a mechanism of the phase transition might also be relevant.  
Further experimental (as well as theoretical) studies on the 
microscopic mechanism of this phase transition are therefore 
highly desirable.  

\section{Summary}

We have studied the electronic states of a vanadate material 
Bi$_x$V$_8$O$_{16}$, a possibly new charge and orbital ordering 
system with the $t_{2g}$-orbitals on a 1D triangular lattice 
with the mixed valency of $3d^2$ : $3d^1$ = 2 : 1.  
By assuming the charge ordering pattern, we have derived the 
effective spin-orbit Hamiltonian by the second-order perturbation 
theory.  Within the approximation neglecting small off-diagonal 
hopping parameters, we have found that the Hamiltonian is 
block-diagonal with vanishing orbital-off-diagonal sectors.  
We then have used a numerical diagonalization technique on small 
clusters of this Hamiltonian and have found that a variety of 
the ground-state phases with different orbital ordering patterns 
appear in the parameter space.  
We have argued that the orbital ordering pattern possibly realized 
in experiment should be a state of the partial singlets where 
the nearest-neighbor spin $S=1$ pairs form singlets with leaving 
the nearly free S=1/2 spins.  

Although experimental data on this material are quite limited at 
present, we hope that the present theoretical study will encourage 
further experimental studies to clarify the nature of the charge, 
orbital, and spin degrees of freedom of this interesting material.  

\section*{Acknowledgements}

We would like to thank Prof. K. Yoshimura for enlightening 
discussion on the experimental aspects of Bi$_x$V$_8$O$_{16}$.  
This work was supported in part by Grants-in-Aid for 
Scientific Research (Nos.~11640335 and 12046216) from the 
Ministry of Education, Science, Sports, and Culture of Japan.  
Computations were carried out at the computer centers of 
the Institute for Molecular Science, Okazaki, and the 
Institute for Solid State Physics, University of Tokyo.

\end{document}